\documentclass[reprint,prd,onecolumn,longbibliography,showpacs,notitlepage]{revtex4-1}

\usepackage{mathtools}
\usepackage[utf8]{inputenc}
\usepackage[T1]{fontenc}

\usepackage{tikz}

\usepackage[colorlinks=true,pdftoolbar = true, pdfstartview = FitH, pdfmenubar = true]{hyperref}
\hypersetup{pdftitle=1802.00492 Nonmetricity formulation of general relativity and its scalar-tensor extension,pdfauthor=Laur J\"arv and Mihkel R\"unkla and Margus Saal and Ott Vilson}

\usepackage[all]{hypcap}

\newcommand{\CLC}[3]{ \left\lbrace {}^{#1}{}_{#2 #3} \right\rbrace}

\newcommand{\RLC}{ \overset{ \scalebox{.7}{$\mathclap[\scriptscriptstyle]{\mathrm{LC}}$} }{R}{}}
\newcommand{\RW}{ \overset{ \scalebox{.7}{$\mathclap[\scriptscriptstyle]{\mathrm{W}}$} }{R}{}}
\newcommand{\RSTP}{ \overset{ \scalebox{.7}{\mathclap[\scriptscriptstyle]{\mathrm{STP}}} }{R}{}}

\newcommand{\TLC}{ \overset{ \scalebox{.7}{$\mathclap[\scriptscriptstyle]{\mathrm{LC}}$} }{T}{}}
\newcommand{\TW}{ \overset{ \scalebox{.7}{$\mathclap[\scriptscriptstyle]{\mathrm{W}}$} }{T}{}}
\newcommand{\TSTP}{ \overset{ \scalebox{.7}{$\mathclap[\scriptscriptstyle]{\mathrm{STP}}$} }{T}{}}

\newcommand{\QLC}{ \overset{ \scalebox{.7}{$\mathclap[\scriptscriptstyle]{\mathrm{LC}}$} }{Q}{}}
\newcommand{\QW}{ \overset{ \scalebox{.7}{$\mathclap[\scriptscriptstyle]{\mathrm{W}}$} }{Q}{}}
\newcommand{\QSTP}{ \overset{ \scalebox{.7}{$\mathclap[\scriptscriptstyle]{\mathrm{STP}}$} }{Q}{}}
\newcommand{\QtSTP}{ \overset{ \scalebox{.7}{$\mathclap[\scriptscriptstyle]{\mathrm{STP}}$} }{\tilde{Q}}{}}

\newcommand{\LSTP}{ \overset{ \scalebox{.7}{$\mathclap[\scriptscriptstyle]{\mathrm{STP}}$} }{L}{}}
\newcommand{\PSTP}{ \overset{ \scalebox{.7}{$\mathclap[\scriptscriptstyle]{\mathrm{STP}}$} }{P}{}}

\newcommand{\GSTP}{ \overset{ \scalebox{.7}{$\mathclap[\scriptscriptstyle]{\mathrm{STP}}$} }{\Gamma}{}}

\newcommand{\DLC}{ \overset{ \scalebox{.7}{$\mathclap[\scriptscriptstyle]{\mathrm{LC}}$} }{\nabla}{}}
\newcommand{\DW}{ \overset{ \scalebox{.7}{$\mathclap[\scriptscriptstyle]{\mathrm{W}}$} }{\nabla}{}}
\newcommand{\DSTP}{ \overset{ \scalebox{.7}{$\mathclap[\scriptscriptstyle]{\mathrm{STP}}$} }{\nabla}{}}

\begin{document}

\title{Nonmetricity formulation of general relativity and its scalar-tensor extension}

\author{Laur Järv}
\email{laur.jarv@ut.ee}

\author{Mihkel Rünkla}
\email{mrynkla@ut.ee}

\author{Margus Saal}
\email{margus.saal@ut.ee}

\author{Ott Vilson}
\email{ovilson@ut.ee}

\affiliation{Laboratory of Theoretical Physics, Institute of Physics, University of Tartu, W. Ostwaldi 1, 50411, Tartu, Estonia}



\begin{abstract}
Einstein’s celebrated theory of gravitation can be presented in three forms: general relativity, teleparallel gravity, and the rarely considered before symmetric teleparallel gravity. Extending the latter, we introduce a new class of theories where a scalar field is coupled nonminimally to nonmetricity $Q$, which here encodes the gravitational effects like curvature $R$ in general relativity or torsion $T$ in teleparallel gravity. We point out the similarities and differences with analogous scalar-curvature and scalar-torsion theories by discussing the field equations, role of connection, conformal transformations, relation to $f(Q)$ theory, and cosmology. The equations for spatially flat universe coincide with those of teleparallel dark energy,  thus allowing to explain accelerating expansion.
\end{abstract}


\maketitle

\twocolumngrid


\section{Introduction}\label{sec:intro}

General relativity (GR) assumes Levi-Civita connection and hence implies zero torsion and nonmetricity. GR has a well researched alternative formulation known as teleparallel equivalent of general relativity (TEGR) \cite{Aldrovandi:2013wha} which instead utilizes Weitzenböck connection and elicits vanishing curvature and nonmetricity. However, there exists also a third possibility, to adopt a connection with vanishing curvature and torsion, which provides a basis  for yet another equivalent formulation of GR, the so-called symmetric teleparallel equivalent of general relativity (STEGR) \cite{Nester:1998mp}, hardly ever studied in the literature \cite{Adak:2005cd,Adak:2008gd,Adak:2004uh}. Instead of curvature $R$, or torsion $T$, it relies on the nonmetricity $Q$ to describe the effects of gravity.

Although TEGR is considered to be completely equivalent to GR, some features make it appealing to study, e.g., the gauge theory structure, possibility to separate inertial and gravitational effects, etc.\ \cite{Aldrovandi:2013wha}. However, interest in this formulation only surged some years ago when it was realized that extensions of TEGR, like $f(T)$ and scalar-torsion gravity differ from their $f(R)$ and scalar-curvature counterparts which extend general relativity \cite{Ferraro:2008ey,Linder:2010py}. Suddenly a completely unexplored new alley opened up for researchers to address the puzzles of dark energy, inflation, etc., resulting in a lot of activity (see Ref.~\cite{Cai:2015emx}). A confusion concerning the local Lorentz invariance has just been recently overcome by stressing the covariant formulation \cite{Krssak:2015oua} and deriving the appropriate equation for the connection \cite{Golovnev:2017dox, Krssak:2017nlv,Hohmann:2017duq,Hohmann:2018rwf}.

While STEGR also promises a set of nice features \cite{Nester:1998mp,Koivisto:2018aip}, its extensions 
potentially offer yet another totally uncharted territory to map and study. The first pioneering works have just appeared looking at $f(Q)$ theories \cite{BeltranJimenez:2017tkd} and higher derivative generalizations \cite{Conroy:2017yln}. In this paper we propose an extension of STEGR by introducing a scalar field that is nonminimally coupled to the nonmetricity scalar $Q$. Our setup resembles the generic forms of scalar-curvature~\cite{EspositoFarese:2000ij,Flanagan:2004bz,Jarv:2014hma} and scalar-torsion~\cite{Geng:2011aj,Jarv:2015odu} theories, where the scalar field is coupled to the curvature and torsion scalar, respectively. Nonminimal couplings arise naturally when quantum effects for a minimal scalar are considered in GR \cite{Birrell:1982ix}, and are utilized in, e.g., the Higgs inflation \cite{Bezrukov:2007ep}.

Let us clarify from the outset that STEGR differs from typical metric-affine theories of gravity \cite{Hehl:1976my,Hehl:1994ue} where GR is extended by allowing connection to possess besides curvature also torsion and nonmetricity, whereby one usually needs specific types or properties of matter to excite and probe such additional geometric structures \cite{Obukhov:1993pt,Neeman:1996zcr,Obukhov:2004hv,Puetzfeld:2007hr,Vitagliano,Ariki:2017qov,Latorre:2017uve}. By imposing vanishing curvature and torsion, in STEGR the GR gravitational action is rewritten in terms of nonmetricity, and all gravitational effects that are attributed to curvature in GR, now equivalently stem from nonmetricity. Therefore in STEGR the matter content can remain unaltered, for in analogy to the Einstein's equations it is the usual matter energy-momentum that is the source of nonmetricity. In our construction novel features appear when a scalar field is nonminimally coupled.

We begin in Sec.~\ref{s:ConnectionsGravities}
with a basic introduction to the key geometric notions and establish the equivalence of general relativity to teleparallel and symmetric teleparallel theories. Next in Sec.~\ref{s:ActionFieldEq} 
we postulate the action, derive the field equations and comment their main features. Then in Sec.~\ref{s:FurtherRemarks} 
we probe the conformal transformations and also show how $f(Q)$ theories fit into the picture. Finally, Sec.~\ref{s:FLRW} 
briefly look at the cosmological equations for spatially flat spacetime, and Sec.~\ref{s:Conclusions} concludes the paper.


\section{Connections, geometries, and gravitational theories}
\label{s:ConnectionsGravities}


\subsection{Decomposition of affine connection}
\label{ss:RelationRTQ}

On metric-affine spacetimes the metric $g_{\mu\nu}$ encodes distances and angles, while the connection $\Gamma^{\lambda}{}_{\sigma\rho}$ independently defines parallel transport and covariant derivatives, e.g.,
\begin{equation}
\label{Covariant derivative}
\nabla_\mu \mathcal{T}^\lambda{}_\nu =  \partial_\mu \mathcal{T}^\lambda{}_\nu  + \Gamma^\lambda{}_{\mu \alpha} \mathcal{T}^\alpha{}_\nu  -  \Gamma^\alpha{}_{\mu \nu} \mathcal{T}^\lambda{}_\alpha \,.
\end{equation}
As known from differential geometry (see, e.g., \cite{Hehl:1994ue,Ortin:2015hya}), generic affine connection can be decomposed into three parts,
\begin{equation}
\label{Connection decomposition}
\Gamma^{\lambda}_{\phantom{\alpha}\mu\nu} =
\left\lbrace {}^{\lambda}_{\phantom{\alpha}\mu\nu} \right\rbrace +
K^{\lambda}_{\phantom{\alpha}\mu\nu}+
 L^{\lambda}_{\phantom{\alpha}\mu\nu} \,,
\end{equation}
viz., the Levi-Civita connection of the metric $g_{\mu\nu}$,
\begin{equation}
\label{LeviCivita}
 \CLC{\lambda}{\mu}{\nu} \equiv \frac{1}{2} g^{\lambda \beta} \left( \partial_{\mu} g_{\beta\nu} + \partial_{\nu} g_{\beta\mu} - \partial_{\beta} g_{\mu\nu} \right) \,,
\end{equation}
contortion
\begin{equation}
\label{Contortion}
K^{\lambda}{}_{\mu\nu} \equiv \frac{1}{2} g^{\lambda \beta} \left( T_{\mu\beta\nu}+T_{\nu\beta\mu} +T_{\beta\mu\nu} \right) = - K_{\nu\mu}{}^{\lambda}\, ,
\end{equation}
and disformation
\begin{equation}
\label{Disformation}
L^{\lambda}{}_{\mu\nu} \equiv \frac{1}{2} g^{\lambda \beta} \left( -Q_{\mu \beta\nu}-Q_{\nu \beta\mu}+Q_{\beta \mu \nu} \right) = L^{\lambda}{}_{\nu\mu}  \,.
\end{equation}
The last two quantities are defined via torsion
\begin{equation}
\label{TorsionTensor}
T^{\lambda}{}_{\mu\nu}\equiv \Gamma^{\lambda}{}_{\mu\nu}-\Gamma^{\lambda}{}_{\nu\mu}\,
\end{equation}
and nonmetricity
\begin{equation}
\label{NonMetricityTensor}
Q_{\rho \mu \nu} \equiv \nabla_{\rho} g_{\mu\nu} = \partial_\rho g_{\mu\nu} - \Gamma^\beta{}_{\rho \mu} g_{\beta \nu} -  \Gamma^\beta{}_{\rho \nu} g_{\mu \beta}  \,.
\end{equation}

\begin{figure}
	\begin{tikzpicture}
	\tikzset{venn circle/.style={draw,circle,minimum width=5cm,fill=#1,opacity=0.1}}
	
	\node [venn circle = red] (A) at (0,0) {};
	\node [venn circle = blue] (B) at (0:2.5cm) {};
	\node [venn circle = green] (C) at (-60:2.5cm) {};
	\node[above] at (barycentric cs:A=-2.7,B=1/2,C=1/2) {Riemann-Cartan}; 
	\node[below] at (barycentric cs:A=-2.7,B=1/2,C=1/2) {$\scriptstyle{Q_{\rho\mu\nu}=0}$}; 
	\node[above] at (barycentric cs:B=-2.7,A=1/2,C=1/2) {torsion free}; 
	\node[below] at (barycentric cs:B=-2.7,A=1/2,C=1/2) {$\scriptstyle{T^\lambda{}_{\mu\nu}=0}$}; 
	\node[above] at (barycentric cs:C=-2.7,A=1/2,B=1/2) {teleparallel}; 
	\node[below] at (barycentric cs:C=-2.7,A=1/2,B=1/2) {$\scriptstyle{R^\sigma{}_{\rho\mu\nu}=0}$};
	\node at (barycentric cs:A=1,B=1,C=-0.6) {\begin{tabular}{c} Riemann \\$\scriptstyle{\QLC_{\rho\mu\nu}=0}$,\\ $\scriptstyle{\TLC^\lambda{}_{\mu\nu}=0}$ \end{tabular}};
	\node at (barycentric cs:A=1,C=1,B=-2/3) {\begin{tabular}{c}Weitzenb\"ock \\ $\scriptstyle{\QW_{\rho\mu\nu}=0,}$ \\ $\scriptstyle{\RW^\sigma{}_{\rho\mu\nu}=0}$ \end{tabular}}; 
	\node at (barycentric cs:B=1,C=1,A=-2/3) {\begin{tabular}{c} symmetric \\ teleparallel \\ $\scriptstyle{\RSTP^\sigma{}_{\rho\mu\nu}=0,}$ \\ $\scriptstyle{\TSTP^\lambda{}_{\mu\nu}=0}$\end{tabular}}; 
	\node[above] at (barycentric cs:A=1/3,B=1/3,C=1/3 ){Minkowski};
	
	\end{tikzpicture}  
	\caption{Subclasses of metric-affine geometry, depending on the properties of connection.}
	\label{fig:geometries}
\end{figure}
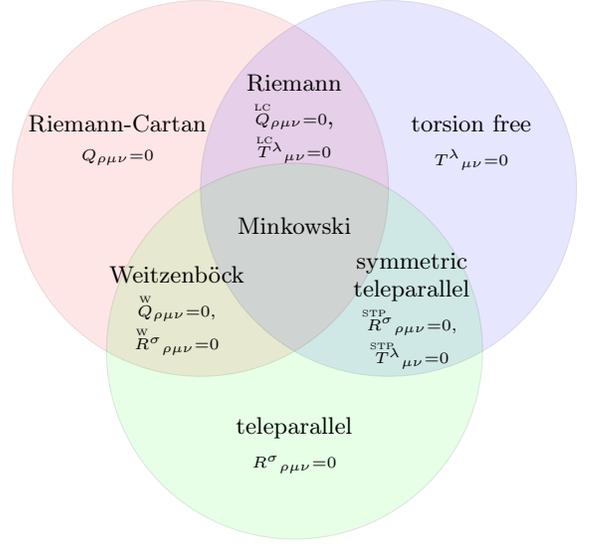

Note that torsion, nonmetricity, as well as curvature
\begin{equation}
\label{Riemann_tensor}
R^{\sigma}{}_{\rho\mu\nu} \equiv \partial_{\mu} \Gamma^{\sigma}{}_{\nu\rho} - \partial_{\nu} \Gamma^{\sigma}{}_{\mu\rho} + \Gamma^{\alpha}{}_{\nu\rho} \Gamma^{\sigma}{}_{\mu\alpha} - \Gamma^{\alpha}{}_{\mu\rho} \Gamma^{\sigma}{}_{\nu\alpha} \,
\end{equation}
are strictly speaking all properties of the connection. By making assumptions about the connection we restrict the generic metric-affine geometry, see Fig.~\ref{fig:geometries}. Taking nonmetricity to vanish gives Riemann-Cartan geometry, taking curvature to vanish gives teleparallel geometry (since the parallel transport of vectors becomes independent of the path), while taking torsion to vanish is just known as torsion free geometry.
We can also impose double conditions on the connection. Vanishing torsion and nonmetricity leaves us with Levi-Civita (LC) connection and Riemann geometry. Assuming nonmetricity and curvature to be zero is the premise of Weitzenböck (W) connection. Keeping torsion and curvature to zero means symmetric teleparallel (STP) connection and geometry. Finally, setting all three to zero yields Minkowski space.
To denote a situation where a particular property is imposed on the connection, and consequently on the covariant derivative, curvature, etc., we use overset labels, e.g., $\GSTP{}^\lambda{}_{\mu\nu}$, $\DW_\mu$,  $\RLC^{\sigma}{}_{\rho\mu\nu}$.


\subsection{Three equivalent formulations of Einstein's gravity}

In order to define a theory of gravity we need to fix the underlying geometry as well as the quantity standing in the action. In laying the grounds for GR Einstein chose Levi-Civita connection, and since nonmetricity and torsion vanish, it remains the task of curvature to encode gravitational dynamics. A suitably constructed invariant quantity of curvature, the curvature scalar,
\begin{equation}
\label{Ricci_scalar}
R \equiv g^{\nu\rho} R^{\mu}{}_{\rho\mu\nu} \,,
\end{equation} 
together with the Levi-Civita provision establishes the Lagrangian for the theory. Both ingredients are important, since giving the action with the curvature scalar, but making only the assumption of vanishing nonmetricity (allowing both nontrivial curvature and torsion) gives a different theory with extra features, Einstein–Cartan–Sciama–Kibble gravity \cite{Kibble:1961ba, 1962rdgr.book..415S}.

It is remarkable that an alternative set of assumptions can yield a theory equivalent to GR. To witness it, let us first rewrite the generic curvature tensor \eqref{Riemann_tensor} as (c.f.\ \cite{Aldrovandi:2013wha,Penas:2017ltd})
\begin{eqnarray}
\label{Riemann_tensor_splitted_LC}
R^{\sigma}_{\phantom{\alpha}\rho\mu\nu} &=& \RLC^{\sigma}_{\phantom{\alpha}\rho\mu\nu} + \DLC_{\mu}  M^{\sigma}_{\phantom{\alpha}\nu\rho} - \DLC_{\nu}  M^{\sigma}_{\phantom{\alpha}\mu\rho}  \nonumber \\
&&+ M^{\alpha}_{\phantom{\alpha}\nu\rho} M^{\sigma}_{\phantom{\alpha}\mu\alpha} - M^{\alpha}_{\phantom{\alpha}\mu\rho} M^{\sigma}_{\phantom{\alpha}\nu\alpha} \,,
\end{eqnarray}
where we used the decomposition \eqref{Connection decomposition} to separate the Levi-Civita terms from the contortion and disformation contributions, collectively denoted as
\begin{equation}
M^{\lambda}_{\phantom{\alpha}\mu\nu}=
K^{\lambda}_{\phantom{\alpha}\mu\nu}+
 L^{\lambda}_{\phantom{\alpha}\mu\nu} \, .
\end{equation}
Now contracting the curvature tensor \eqref{Riemann_tensor_splitted_LC} to form the curvature scalar \eqref{Ricci_scalar} yields
\begin{eqnarray}
\label{contracted_Riemann_for_LC}
R &=&  \RLC + M^{\alpha}_{\phantom{\alpha}\nu\rho} M^{\mu}_{\phantom{\alpha}\mu\alpha} g^{\nu\rho} - M^{\alpha}_{\phantom{\alpha}\mu\rho} M^{\mu}_{\phantom{\alpha}\nu\alpha} g^{\nu\rho} 
\nonumber \\
&&+ \DLC_{\mu} \left( M^{\mu}_{\phantom{\alpha}\nu\rho} g^{\nu\rho} - M^{\nu}_{\phantom{\alpha}\nu\rho} g^{\mu\rho} \right) \,.
\end{eqnarray}
It is obvious, that if we restrict the geometry to have vanishing  torsion and nonmetricity, the curvature scalar (\ref{contracted_Riemann_for_LC}) is simply
\begin{equation}
\label{RinGR}
R  = \RLC \,.
\end{equation}
This is the case of general relativity. 

If we instead choose to work in the setting of Weitzenböck connection whereby the curvature and the nonmetricity are zero, then Eq.\ (\ref{contracted_Riemann_for_LC}) yields
\begin{equation}
\label{RandT}
\RLC =-\TW-2\DLC_{\alpha} \TW^{\alpha} \,.
\end{equation}
Here we introduced the torsion scalar, defined in principle for arbitrary connection as
\begin{equation}
\label{TorsionScalar}
T\equiv\frac{1}{4}T_{\alpha \beta \gamma} T^{\alpha \beta \gamma} + \frac{1}{2} T_{\alpha \beta \gamma}T^{\gamma \beta \alpha} - T_{\alpha}T^{\alpha} \,,
\end{equation}
and the one independent contraction of the torsion tensor,
\begin{equation}
\label{TorsionContractions}
T_{\mu}\equiv T^{ \alpha}{}_{\mu\alpha}=-T^{\alpha}{}_{\alpha\mu } \,.
\end{equation}
The Weitzenböck torsion scalar in Eq.~\eqref{RandT} differs from the GR curvature scalar by a total divergence term. Therefore a theory where the action is set by the torsion scalar (\ref{TorsionScalar}), restricted to Weitzenböck connection, should give equivalent field equations to GR. This is indeed the case, known as teleparallel equivalent of general relativity \cite{Aldrovandi:2013wha}.
 
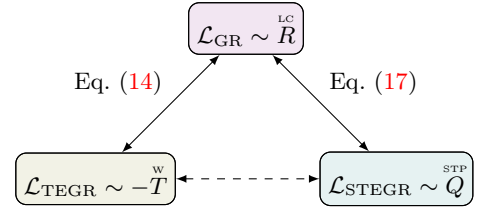
\begin{figure}
	\tikzstyle{block} = [rectangle, draw, text centered, rounded corners]
	\begin{tikzpicture}
	\node [block,fill={rgb:red,1;blue,1},fill opacity=0.1,text opacity=0] (a) at (0,0) {$\mathcal{L}_{\mathrm{GR}} \sim \RLC$};
	\node [block] at (0,0) {$\mathcal{L}_{\mathrm{GR}} \sim \RLC$};
	\node [block,fill={rgb:red,1;green,1},fill opacity=0.1,text opacity=0] (b) at (-2,-2) {$\mathcal{L}_{\mathrm{TEGR}} \sim - \TW $};
	\node [block] (b) at (-2,-2) {$\mathcal{L}_{\mathrm{TEGR}} \sim - \TW $};
	\node [block,fill={rgb:blue,1;green,1},fill opacity=0.1,text opacity=0] (c) at (2,-2)  {$ \mathcal{L}_{\mathrm{STEGR}} \sim \QSTP $};
	\node [block] (c) at (2,-2)  {$ \mathcal{L}_{\mathrm{STEGR}} \sim \QSTP $};
	\draw [<->,>=latex](a)  to node [above left] {Eq.\ (\ref{RandT})}  (b);
	\draw [<->,>=latex](a)  to node [above right] {Eq.\ (\ref{RandQ})} (c);
	\draw [<->,>=latex,dashed](b)  -- (c);
	\end{tikzpicture}  
	\caption{Triple equivalence of gravitational theories: general relativity (GR) based on Levi-Civita connection with vanishing nonmetricity and torsion, teleparallel equivalent of general relativity (TEGR) based on Weitzenböck connection with vanishing nonmetricity and curvature, and symmetric teleparallel equivalent of general relativity (STEGR) based on connection with vanishing curvature and torsion.}
	\label{fig:theories}
\end{figure}

A third possibility, hardly explored before, is to impose vanishing curvature and torsion, which is the case of symmetric teleparallel connection. Plugging $\RSTP^{\sigma}{}_{\rho\mu\nu}=0$ and $M^{\lambda}_{\phantom{\alpha}\mu\nu}= \LSTP^{\lambda}_{\phantom{\alpha}\mu\nu}$ into (\ref{contracted_Riemann_for_LC}) now yields
\begin{equation}
\label{RandQ}
\RLC=\QSTP- \DLC_{\alpha}(\QSTP^{\alpha}-\QtSTP^{\alpha}) \,.
\end{equation}
Here the nonmetricity scalar is defined for arbitrary connection as 
\begin{equation}
\label{NonmetricityScalar}
Q=-\frac{1}{4}Q_{\alpha \beta \gamma}Q^{\alpha \beta \gamma}+\frac{1}{2}Q_{\alpha \beta \gamma}Q^{ \gamma \beta \alpha}+\frac{1}{4}Q_{\alpha}Q^{\alpha}-\frac{1}{2}Q_{\alpha}\tilde{Q}^{\alpha} \,,
\end{equation}
while the nonmetricity tensor is endowed with two independent contractions,
\begin{equation}
\label{NonMetricityContractions}
Q_{\mu}\equiv Q_{\mu \ \alpha}^{\ \: \alpha} \, , \qquad \tilde{Q}^{\mu} \equiv Q_{\alpha }^{\ \: \alpha \mu}  \, .
\end{equation}
As the action constructed with the nonmetricity scalar (\ref{NonmetricityScalar}), restricted to symmetric teleparallel connection, would differ from the GR action only by a total divergence term, and the latter does not contribute to the equations of motion, we get another formulation of Einstein's gravity, symmetric teleparallel equivalent of general relativity \cite{Nester:1998mp,Adak:2005cd,Adak:2008gd,Adak:2004uh,BeltranJimenez:2017tkd,Koivisto:2018aip}. All three equivalent formulations are summarized by Fig.~\ref{fig:theories}.


\section{Scalar-nonmetricity theory}
\label{s:ActionFieldEq}


\subsection{Action}
\label{ss:Action}

Let us take the metric $g_{\mu\nu}$, the connection $\Gamma^{\lambda}{}_{\sigma\rho}$, and the scalar field $\Phi$ as independent variables and consider the action functional
\begin{equation}
\label{Action}
S = \frac{1}{2} \int\mathrm{d}^4x \sqrt{-g} \left( \mathcal{L}_\mathrm{g} +  \mathcal{L}_{\ell} \right) + S_{\mathrm{m}}\,,
\end{equation}
with gravitational Lagrangian
\begin{eqnarray}
\label{L_g}
\mathcal{L}_\mathrm{g} &=& \mathcal A(\Phi) Q -\mathcal B(\Phi) g^{\alpha\beta}\partial_\alpha\Phi \partial_\beta\Phi 
-2{\mathcal V}(\Phi) \,
\end{eqnarray}
and Lagrange multipliers terms
\begin{eqnarray}
\label{L_l}
\mathcal{L}_{\ell} &=& 2\lambda_{\mu}^{\ \beta \alpha \gamma  } R^{\mu}_{\ \beta \alpha \gamma  } + 2\lambda_{\mu}^{\  \alpha \beta} T^{\mu}_{\ \ \alpha \beta} \,,
\end{eqnarray}
while $S_{\mathrm{m}} = S_{\mathrm{m}}\left[ g_{\mu\nu}, \chi \right]$ denotes the action of matter fields $\chi$ which are not directly coupled to the scalar field $\Phi$. 

In analogy with the scalar-curvature~\cite{Flanagan:2004bz,Jarv:2014hma} and scalar-torsion~\cite{Jarv:2015odu,Hohmann:2018rwf} theories, $\mathcal {A}(\Phi)$, $\mathcal{B}(\Phi)$ and $\mathcal{V}(\Phi)$ are functions. The nonmetricity scalar $Q$ is given by (\ref{NonmetricityScalar}). In the case when $\mathcal{A}=1$ and $\mathcal{B}=\mathcal{V}=0$ the theory reduces to plain STEGR. The Lagrange multipliers, assumed to respect the antisymmetries of the associated geometrical objects, i.e., $\lambda_{\mu}^{\ \beta \alpha \gamma  } = \lambda_{\mu}^{\ \beta [\alpha \gamma] }$ and $\lambda_{\mu}^{\  \alpha \beta}=\lambda_{\mu}^{\  [\alpha \beta]}$, impose vanishing curvature $R^{\mu}_{\ \beta \alpha \gamma  }=0$ and torsion $T^{\mu}_{\ \alpha \beta}=0$, as expected in the symmetric teleparallel framework.


\subsection{Field equations for metric and scalar field}
\label{ss:FieldEquations}

Varying the action \eqref{Action} with respect to the metric, and keeping in mind that the connection is flat and torsion-free, yields
\begin{eqnarray}
\nonumber
\mathcal{T}_{\mu\nu} &=&
\frac{2}{\sqrt{-g}}\DSTP_{\alpha} \left( \sqrt{-g}\mathcal{A} \PSTP^{\alpha}_{\hphantom{\lambda}\mu\nu} \right)  - \frac{1}{2} g_{\mu\nu} \mathcal A \QSTP {} 
\\
\nonumber
&& +  \mathcal{A} \left( \PSTP_{\mu\alpha\beta} \QSTP_{\nu}^{\hphantom{\nu} \alpha\beta} - 2\QSTP_{\alpha\beta\mu} \PSTP^{\alpha\beta}_{\hphantom{\alpha\beta} \nu} \right) + \\
\label{MetricFieldEq}
&& + \frac{1}{2} g_{\mu\nu} \left( \mathcal B g^{\alpha\beta}\partial_\alpha\Phi \partial_{\beta} \Phi + 2 {\mathcal V}  \right) - \mathcal{B} \partial_{\mu} \Phi \partial_\nu \Phi \,, \quad
\end{eqnarray}
where we introduced the nonmetricity conjugate (or superpotential) \cite{BeltranJimenez:2017tkd}
\begin{eqnarray}
\nonumber
\PSTP^{\alpha}_{\ \mu \nu} &\equiv& -\frac{1}{2} \LSTP^{\alpha}_{\ \mu \nu} + \frac{1}{4} \left( \QSTP^{\alpha} - \QtSTP^{\alpha} \right) g_{\mu \nu} \\
\label{NMconjugate}
&&{} - \frac{1}{8} \left( \delta^{ \alpha }_{\mu} \QSTP^{\phantom{\alpha}}_{\nu} + \delta^{\alpha }_{\nu} \QSTP^{\phantom{\alpha}}_{\mu} \right)
  \,, \quad
\end{eqnarray}
which satisfies $\QSTP= \QSTP_{\alpha}^{\hphantom{\alpha} \mu \nu} \PSTP^{\alpha}_{\hphantom{\alpha} \mu \nu}$.
The usual matter energy-momentum tensor, 
\begin{eqnarray}
\label{EnergyMomentum}
\mathcal{T}_{\mu \nu} \equiv -\frac{2}{\sqrt{-g}} \frac{\delta S_{\mathrm{m}}\left[ g_{\sigma\rho}, \chi \right] }{\delta g^{\mu \nu}} \,,
\end{eqnarray}
acts as a source to gravity which is described by nonmetricity. Note that for minimal coupling, $\mathcal{A}=1$, the two first lines of \eqref{MetricFieldEq} are in fact Einstein's equations of GR, since
we can write
\begin{eqnarray}
\RLC_{\mu \nu} - \frac{1}{2} g_{\mu \nu} \RLC &=& \frac{2}{\sqrt{-g}}\DSTP_{\alpha} \left( \sqrt{-g} \PSTP^{\alpha}_{\hphantom{\lambda}\mu\nu} \right)  - \frac{1}{2} g_{\mu\nu} \QSTP {} \nonumber \\
&&
+  \PSTP_{\mu\alpha\beta} \QSTP_{\nu}^{\hphantom{\nu} \alpha\beta} - 2\QSTP_{\alpha\beta\mu} \PSTP^{\alpha\beta}_{\hphantom{\alpha\beta} \nu} 
\end{eqnarray}
from Eq.~\eqref{Riemann_tensor_splitted_LC}.

Variation with respect to the scalar field yields
\begin{eqnarray}
\label{ScalarFieldEq}
2\mathcal B \DLC_{\alpha}\DLC^{\alpha} \Phi 
+\mathcal {B}^{\prime}g^{\alpha\beta}\partial_{\alpha}\Phi\partial_{\beta}\Phi + \mathcal{A}^{\prime} \QSTP -
2\mathcal{V}^{\prime} =0 \,,
\end{eqnarray}
where the primes mean derivative with respect to the scalar field.
Like in the scalar-curvature and scalar-torsion case, the scalar field equation obtains a term with the geometric invariant to which the scalar is nonminimally coupled to, here the nonmetricity scalar $\QSTP$. In the scalar-curvature case the curvature scalar $\RLC$ contains second derivatives of the metric and it is natural to seek to ``debraid'' \cite{Bettoni:2015wta} the equations by substituting in the trace of the metric field equations, thereby removing $\RLC$ but introducing the trace of matter energy-momentum  in the scalar field equation  \cite{Jarv:2015kga}. In the nonmetricity case the Eqs.\ (\ref{MetricFieldEq}) and (\ref{ScalarFieldEq}) are already debraided. Apparently, the role of matter as a source for the scalar field in scalar-nonmetricity gravity, like in scalar-torsion gravity~\cite{Hohmann:2018rwf}, is more indirect than in scalar-curvature gravity~\cite{Flanagan:2004bz,Jarv:2014hma}.


\subsection{Variation with respect to the connection}
\label{ss:Connection}

Variation of the action \eqref{Action} with respect to connection yields an equation containing Lagrange multipliers,
\begin{eqnarray}
\label{ConnectionEq}
\DSTP_{\gamma} \left(\sqrt{-g} \lambda_{\mu}{}^{\beta \alpha \gamma} \right)+ \sqrt{-g} \lambda_{\mu}{}^{\alpha \beta} =
\sqrt{-g} \mathcal{A} \PSTP^{\alpha \beta}{}_{\mu} \,.
\end{eqnarray}
Due to the vanishing curvature and torsion the covariant derivatives commute, hence one can eliminate the Lagrange multipliers, which are antisymmetric with respect to their last indices, from (\ref{ConnectionEq}) by acting on it with $\DSTP_{\beta}\DSTP_{\alpha}$. This yields the following equations
\begin{eqnarray}
\label{ConnectionEq1}
\DSTP_{\beta} \DSTP_{\alpha}(\sqrt{-g} \mathcal{A} \PSTP^{\alpha \beta}{}_{\mu})
&=& 0 \,,
\end{eqnarray}
which can be simplified further by using the equivalent of Bianchi identity, $\DSTP_\beta \DSTP_\alpha (\sqrt{-g} \PSTP^{\alpha\beta}{}_\mu)=0$, to give
\begin{align}
\label{ConnectionEq2}
\DSTP_\beta \left\lbrace \partial_\alpha \mathcal{A} \left[ \DSTP_\mu (\sqrt{-g} g^{ \alpha \beta}) - \delta^{\alpha}_{\mu} \DSTP_\gamma(\sqrt{-g} g^{\gamma\beta }) \right]\right\rbrace = 0 \,. 
\end{align}

Thus, variation with respect to the connection gave us four equations. These can be understood as follows. If one demands curvature and torsion to vanish, then there exists a particular coordinate system where all symmetric teleparallel connection coefficients vanish \cite{Bao:2000,Schrodinger:1985}, a configuration called coincident gauge \cite{BeltranJimenez:2017tkd}. Therefore generic symmetric teleparallel connection in an arbitrary coordinate system can be obtained by a coordinate transformation from the coincident gauge and represented as
\begin{equation}
\label{STP general connection}
\GSTP^{\lambda}_{\ \mu \nu}=\frac{\partial x^{\lambda}}{\partial \xi^{\alpha}} \left( \frac{\partial}{\partial x^{\mu}} \frac{\partial \xi^{\alpha}}{\partial x^{\nu}} \right) \,,
\end{equation}
where $\xi^{\alpha}$ are some functions. The connection equations \eqref{ConnectionEq1} fix the four freedoms encoded by $\xi^{\alpha}$, and guarantee that the connection coefficients are consistent with the chosen metric. In fact, if we had assumed from the beginning that the connection is of the form \eqref{STP general connection}, then the variation of the action \eqref{Action} with respect to $\xi^\alpha$ would have given the same equations \eqref{ConnectionEq1}. These equations are first order differential equations for the connection, and second order for the Jacobian matrix $\frac{\partial \xi^{\alpha}}{\partial x^{\nu}} $.

This state of affairs can be compared to the scalar-torsion gravities with Weitzenböck connection. There demanding vanishing nonmetricity and curvature is not able to set the connection coefficients to zero in some coordinate basis, but it is nevertheless possible in a noncoordinate basis, i.e., in some frame. Generic Weitzenböck connection is thus generated not by coordinate transformations but by local Lorentz transformations from this frame \cite{Krssak:2015oua}. Lorentz transformations have six independent parameters, resulting in six freedoms in the connection, which are then fixed by the six equations coming from the variation of the action with respect to flat and nonmetricity free connection \cite{Golovnev:2017dox, Krssak:2017nlv, Hohmann:2017duq, Hohmann:2018rwf}. These equations are first order for the connection, and second order for the Lorentz matrix.

Finally, let us note that in the pure STEGR case with $\mathcal{A}=1$ the equation \eqref{ConnectionEq1} reduces to the Bianchi identity and the symmetric teleparallel connection is not present in the equations for the metric and (in this case) minimally coupled scalar field. This is again like in the TEGR case \cite{Aldrovandi:2013wha}, whereby one may still entertain other types of arguments to restrict the connection \cite{Krssak:2015rqa,Krssak:2015lba}.


\subsection{Conservation of matter energy-momentum}
Taking the the Levi-Civita covariant divergence of the field equations (\ref{MetricFieldEq}), and using the scalar field equation (\ref{ScalarFieldEq}) as well as the connection equation \eqref{ConnectionEq1} one can derive the continuity equation for matter fields,
\begin{equation}
\label{ContinuityEquation}
\DLC_{\alpha} \mathcal{T}^{\alpha}_{\phantom{\alpha} \mu} = 0 \,.
\end{equation}
This equation also follows from the diffeomorphism invariance of the matter action \cite{Koivisto:2005yk}.

We conclude that there are three independent equations out of (\ref{MetricFieldEq}), (\ref{ScalarFieldEq}), (\ref{ConnectionEq1}), (\ref{ContinuityEquation}), quite in analogy with the scalar-torsion case \cite{Hohmann:2018rwf}. If one makes an ansatz for the metric, connection, and the scalar field, one has to check that the ansatz is consistent with this set of equations, including the connection equation. 


\section{Further remarks}
\label{s:FurtherRemarks}


\subsection{Conformal transformations}
\label{ss:ConformalStructure}
 
Contrary to the scalar-curvature case \cite{Flanagan:2004bz,Jarv:2014hma} the action (\ref{Action}) does not preserve its form under the local conformal rescaling of the metric. The nonmetricity scalar transforms under the conformal transformation $\bar{g}_{\mu \nu}=\mathrm{e}^{\Omega(\Phi)}g_{\mu \nu}$ as follows:
\begin{eqnarray}
\label{NonMetricityScalarRescaled}
\bar{Q}=\mathrm{e}^{-\Omega}\Bigl(Q + \frac{3}{2} g^{\alpha\beta} \partial_{\alpha} \Omega\partial_{\beta} \Omega+ (Q^{\alpha}-\tilde{Q}^{\alpha})\partial_{\alpha}\Omega \Bigr) . \quad
\end{eqnarray}
The additional piece proportional to $g^{\alpha\beta}\partial_{\alpha} \Omega\partial_{\beta} \Omega$ can be absorbed into the redefinition of the kinetic term of the scalar field, however the piece $(Q^{\alpha}-\tilde{Q}^{\alpha})\partial_{\alpha}\Omega$ does not appear in the original action. The latter causes the action (\ref{Action}) not to preserve its structure under conformal transformations. 

However, if we add a term $(Q^{\alpha}-\tilde{Q}^{\alpha})\partial_{\alpha}\mathcal{A}(\Phi)$  to the original Lagrangian (\ref{L_g}), we obtain the equivalent to the familiar scalar-curvature theory, which is covariant under the conformal transformations and scalar field redefinitions. Introducing this term, multiplied by a function would give a theory which interpolates between scalar-curvature and scalar-nonmetricity theories.
This is similar to the case of scalar-torsion theories and their generalizations \cite{Bamba:2013jqa, Bahamonde:2015hza, Wright:2016ayu,Hohmann:2018ijr}, where one has to include the boundary term relating the Ricci and torsion scalars in order to obtain a conformally invariant action.


\subsection{Scalar-nonmetricity equivalent of $f(Q)$ theory}
\label{ss:F(Q)}

It is easy to show that $f(Q)$ theories \cite{BeltranJimenez:2017tkd} where
\begin{equation}
\mathcal{L}_{\mathrm{g},f(Q)} = f(Q) \,
\label{lambda_1}
\end{equation}
form a particular subclass of scalar-nonmetricity theories. 
Following the standard procedure used in scalar-curvature \cite{Teyssandier:1983zz, Wands:1993uu, Chiba:2003ir} and scalar-torsion \cite{Izumi:2013dca} cases, let us introduce an auxiliary field $\Phi$ to write
\begin{equation}
\mathcal{L}_{\mathrm{g},aux} = f'(\Phi) \, Q  -  \left(f'(\Phi) \, \Phi - f(\Phi) \right) \,.
\label{lambda_3}
\end{equation}
By varying the action (\ref{lambda_3}) with respect to $\Phi$ 
yields $f''(\Phi) ( \Phi  -  Q) = 0$. Provided $f''(\Phi) \neq 0$ this equation implies $\Phi=Q$ and restores the original Lagrangian (\ref{lambda_1}). With identifications $\mathcal{A}(\Phi)=f'(\Phi)$, $2 \mathcal{V}(\Phi) = f'(\Phi) \, \Phi - f(\Phi)$ this is identical to the scalar-nonmetricity Lagrangian \eqref{L_g}, where $\mathcal{B}(\Phi) = 0$. Note that contrary to the $f(R)$ case, which in the scalar-curvature representation enjoys a dynamical scalar field, the $f(Q)$ as well as $f(T)$ theory are mapped to a version with nondynamical scalar.
  
  
\section{Example: Friedmann cosmology}
\label{s:FLRW}

Let us consider the spatially flat line element
\begin{equation}
\label{FLRWlineElement}
\mathrm{d}s^2=-\mathrm{d}t^2+a(t)^2\delta_{ij}\mathrm{d}x^i\mathrm{d}x^j \,.
\end{equation}
We can try that the zero connection coefficients $\GSTP^{\lambda}_{\ \mu\nu}=0$ satisfy the connection equation \eqref{ConnectionEq2}, and are thus consistent with the metric \eqref{FLRWlineElement}. A direct calculation yields $\QSTP=-6H^2$, where $H=\frac{\dot{a}}{a}$ is the Hubble parameter and the dot denotes the time derivative. The field equations for perfect fluid matter read:
\begin{eqnarray}
\label{FLRWfieldEq1}
H^2 &=& \frac{1}{3\mathcal{A}}\Bigl(\rho+\frac{1}{2}\mathcal{B}\dot{\Phi}^2+\mathcal{V}\Bigr)\, ,
\\
\label{FLRWfieldEq2}
2\dot{H}+3H^2 &=& \frac{1}{\mathcal{A}}\Bigl(-2\mathcal{A}^\prime H \dot{\Phi}-\frac{1}{2}\mathcal{B}\dot{\Phi}^2+\mathcal{V}-p\Bigr)\,.
\end{eqnarray}
Here $\rho$ is the energy density and $p$ is the pressure of the fluid. Using the scalar field equation
\begin{equation}
\label{FLRWscalarFieldEq}
\mathcal{B}\ddot{\Phi}+(3\mathcal{B}H+\frac{1}{2}\dot{\mathcal{B}}) \dot{\Phi}+\mathcal{V}^{\prime}+3\mathcal{A}^{\prime}H^2=0\, ,
\,
\end{equation}
one can verify that the continuity equation
\begin{equation}
\label{FLRWcontinuityEq}
\dot{\rho}=-3H(\rho+p)
\,
\end{equation}
is sustained. 

It is interesting that these cosmological equations match the corresponding equations in the scalar-torsion counterpart, or teleparallel dark energy, where $\QSTP$ is replaced by $-\TW$ in the action \eqref{Action} \cite{Geng:2011aj,Jarv:2015odu}. Therefore like scalar-curvature and scalar-torsion gravities, the scalar-nonmetricity theory can be also used to explain the early and late time accelerated expansion of the universe.
Moreover, following the result of the previous section, we can further infer, that spatially flat cosmologies in $f(Q)$ and $f(T)$ theories are the same. To understand the detailed mapping between the theories definitely calls for further studies.


\section{Conclusion}
\label{s:Conclusions}

It is remarkable that Einstein's centennial theory of gravity accepts three formulations: general relativity based on curvature, teleparallel gravity based on torsion, and symmetric teleparallel gravity based on nonmetricity. Especially the latter has very little known about it. This work endeavors to explore the ground by considering a generic setting where a scalar field is nonminimally coupled to the nonmetricity scalar in the symmetric teleparallel framework. We derived the field equations, and discussed conformal transformation, relation to $f(Q)$ theories, as well as cosmology, comparing those with the corresponding results in scalar-curvature and scalar-torsion theories. Just as in the latter two cases, the scalar-nonmetricity theory manages to explain both early and late time accelerated expansion of the universe.

A lot of research waits ahead, most obviously constructing solutions and clarifying their features, but also understanding the relations between the theories established in different geometric settings. It might be interesting to consider more general extensions of symmetric teleparallel gravity (in analogy to the recent works in teleparallel gravity \cite{Bahamonde:2017wwk, Hohmann:2017duq,Hohmann:2018vle,Hohmann:2018dqh,Hohmann:2018ijr}), in order to survey the landscape of consistent and observationally viable theories. A broader picture where alternative formulations are taken into account, may well offer novel perspectives and insights into the issues that grapple general relativity.

\vspace{-1cm}

\begin{acknowledgments}
The authors thank Manuel Hohmann, Martin Kr\v{s}\v{s}\'ak and Christian Pfeifer for many useful discussions, and Tomi Koivisto for correspondence. The authors gratefully acknowledge the contribution by the anonymous referees, whose comments helped us clarify potentially confusing statements. The work was supported by the Estonian Ministry for Education and Science through the Institutional Research Support Project IUT02-27 and Startup Research Grant PUT790, as well as the European Regional Development Fund through the Center of Excellence TK133 ``The Dark Side of the Universe''.
\end{acknowledgments}


\onecolumngrid

\bibliography{NMSSTEGR_references}

\end{document}